\documentclass[%
 preprint,
 amsmath,amssymb,
 aps,
superscriptaddress
]{revtex4-2}

\usepackage{graphicx}
\usepackage{dcolumn}
\usepackage{bm}

\setcitestyle{super}

\usepackage{ulem}
\usepackage{upgreek}
\usepackage{color}
\usepackage{upgreek}
\usepackage[usenames,dvipsnames]{xcolor}

\begin{document}


\title{
Phase separation and dynamical arrest of protein solutions \\dominated by short-range attractions}

	\author{Jan Hansen}
\affiliation{Heinrich Heine University, Condensed Matter Physics Laboratory, D\"{u}sseldorf, Germany}

	\author{Carolyn J. Moll}
\affiliation{Heinrich Heine University, Condensed Matter Physics Laboratory, D\"{u}sseldorf, Germany}

	\author{Leticia L\'opez Flores}
\affiliation{Instituto de F\'isica ``Manuel Sandoval Vallarta'', Universidad Aut\'onoma de San Luis Potos\'i, Álvaro Obreg\'on 64, 78000 San Luis Potos\'i, San Luis Potos\'i, M\'exico}
\affiliation{Department of Materials Science and Engineering, Northwestern University, Evanston, Illinois 60208, USA}

\author{Ram\'on Casta\~neda-Priego}
\affiliation{Division of Sciences and Engineering, University of Guanajuato, 37150 Le\'on, M\'exico}

	\author{Magdaleno Medina-Noyola}
\affiliation{Instituto de F\'isica ``Manuel Sandoval Vallarta'', Universidad Aut\'onoma de San Luis Potos\'i, Álvaro Obreg\'on 64, 78000 San Luis Potos\'i, San Luis Potos\'i, M\'exico}

	\author{Stefan U. Egelhaaf}
\affiliation{Heinrich Heine University, Condensed Matter Physics Laboratory, D\"{u}sseldorf, Germany}

	\author{Florian Platten}
	\email{florian.platten@hhu.de}
	\affiliation{Heinrich Heine University, Condensed Matter Physics Laboratory, D\"{u}sseldorf, Germany}

\date{\today}

\begin{abstract}
The interplay of phase separation and dynamical arrest can lead to the formation of gels and glasses, which is relevant for such diverse fields as hard and soft condensed matter physics, materials science, food engineering and pharmaceutical industry. 
Here, 
the non-equilibrium states as well as the interactions of globular proteins are analyzed. 
Lysozyme in brine is chosen as a model system with short-range attractions. 
The metastable gas-liquid binodal and the dynamical arrest line as well as the second virial coefficient $B_2$ have been determined for various solution conditions by cloud-point measurements, optical microscopy, centrifugation experiments and light scattering. 
If temperature is expressed in terms of $B_2$, the binodals obtained under various conditions fall onto a master curve, as suggested by the extended law of corresponding states. 
Arrest lines for different salt concentrations overlap within experimental errors, whereas they do not overlap if the temperature axis is replaced by $B_2$. 
This indicates that the binodals are not sensitive to the details of the potential, but can be described by one integral parameter, i.e.\,\,$B_2$, whereas the arrest line appears governed by its attractive part. 
The experimental findings are supported by numerical results of the non-equilibrium self-consistent generalized Langevin equation theory.
\end{abstract}

\maketitle

\newpage

\section{\label{sec:intro}Introduction}

Colloidal suspensions exhibit various kinds of disordered states, such as repulsive and attractive glasses,\cite{Pusey1987,Pham2002} gels,\cite{Lin1989,Trappe2004} and glassy cluster phases.\cite{Sedgwick2004b,Lu2006}
If the particle volume fraction $\phi$ is low, strong inter-particle attractions can induce irreversible aggregation into ramified gels.\cite{Trappe2004,Zaccarelli2007}
At high $\phi$, a repulsive colloidal glass, in which particles are trapped in cages formed by their neighbors, is formed by hard spheres and hence an arrested disordered state can occur without attractive interactions.\cite{Pusey1987}
At intermediate $\phi$, colloids interacting via short-range attractions can undergo a transition from liquid- to solid-like behavior;\cite{Dawson2002,Bergenholtz2003,Cates2004,Foffi2005} 
 the competition between liquid--liquid phase separation (LLPS) and dynamical arrest can lead to arrested spinodal decomposition.\cite{Verduin1995,Verhaegh1997,Manley2005,Cardinaux2007,Charbonneau2007,Lu2008,Conrad2010}
However, this interplay of phase separation and arrest can be affected by many parameters, including the interaction range \cite{Ramakrishnan2002,Sedgwick2004b} and the existence of repulsive interactions\cite{Stradner2004,Gibaud2012}.

Despite recent progress many questions regarding the transition from liquid- to
solid-like states remain elusive.\cite{Zaccarelli2007,Guo2011,Royall2021} 
While gels
and glasses formed at low and high densities, respectively, are relatively well studied, far less is
known about the arrest transition in moderately concentrated suspensions. 
Theoretical and
simulation studies\cite{Dawson2002,Cates2004} indicate that the dynamical arrest line
connects these limiting cases and hence also includes moderately concentrated suspensions.
Yet, it has been localized in relatively few experiments, which in addition provide conflicting results.
The gel or arrest line was found either to follow the large volume fraction branch of the
binodal of Baxter’s sticky particles for colloid-polymer mixtures\cite{Lu2008} or to extend far into the unstable region below the spinodal for proteins with mixed (short-range attractive and long-range repulsive) potentials\cite{Cardinaux2007,Bucciarelli2015}. 
It is still debated which of these scenarios occurs depending on the strength and range of the interactions.\cite{DaVela2020}

Various experimental systems exhibit gel and glassy states, including colloid-polymer mixtures,\cite{Poon2002,Pham2002,Eckert2002,Manley2005,Royall2018}
nanoparticle dispersions,\cite{Verduin1995,Guo2011,Eberle2012,Zhang2017,Harden2018} and protein solutions\cite{Cardinaux2007,Gibaud2009,Foffi2014,DaVela2017b,DaVela2020}.
Protein solutions represent a convenient model system due to their well-defined properties; 
they are monodisperse and typically carry a small number of charges.\cite{Poon2000,Sedgwick2007} Moreover, the length scales of their condensed states are typically on the order of microns and the time scales associated with their phase transition kinetics range from milliseconds to days; as a consequence, phase transitions in protein solutions can be readily accessed by optical microscopy and light scattering techniques\cite{Muschol1995,Broide1996,Wilson2003,Sedgwick2005} as well as X-ray and neutron scattering at very small angles\cite{DaVela2016,Matsarskaia2019,Girelli2021}.

In an acidic buffer solution lysozyme is prone to crystallize or phase separate.\cite{Muschol1997,Sedgwick2005,Hentschel2021}
In these cases, salt concentrations are so high that electrostatic interactions are largely screened\cite{Platten2015,Platten2016} and hence the inter-particle interactions are dominated by short-range attractions.
Under these conditions, the interactions and, as a consequence, the state boundaries can be modulated by additives, such as guanidin hydrochloride (GuHCl) at low, non-denaturing concentrations.\cite{Platten2015b,Hansen2016,Platten2016,Hansen2021b}
Such short-range attractive systems appear well-suited to quantitatively explore and characterize under which conditions phase coexistence and dynamical arrest occur. 
In the present work, for three salt compositions and a broad range of volume fractions, the liquid--liquid phase coexistence line, the arrest line as well as  the second virial coefficient $B_2$ were determined by visual inspection, centrifugation experiments, optical microscopy and light scattering.
$B_2$ quantifies the strength of protein--protein interactions; 
for a simple, spherosymmetric potential $U(r)$ with center-to-center distance $r$, its definition reads
\begin{equation}\label{eq:b}
B_2 = 2 \pi \int_0^\infty \left( 1 - \exp{\left[-\frac{U(r)}{k_\text{B} T} \right]} \right) r^2 \text{d}r \, 
\end{equation}
with thermal energy $k_\text{B}T$.\cite{Ben-Naim,Santos} 
Often, $B_2$ is normalized by the second virial coefficient of a corresponding hard-sphere system with the same diameter $\sigma$, $B_2^\text{HS}=(2\pi/3)\,\sigma^3$, and reported as  $b_2=B_2/B_2^\text{HS}$. 
The temperature axis of the state diagrams can be expressed in terms of $b_2$.
Our experimental results are compared with numerical results of the non-equilibrium self-consistent generalized Langevin equation (NE-SCGLE) theory.
This work hence aims at a more comprehensive and conclusive description of phase separation and dynamical arrest in systems dominated by short-range attractions.

\section{\label{sec:mm}Materials and Methods}

\subsection{Sample preparation}
Hen egg-white lysozyme powder (Sigma Aldrich, prod.~no.~L6876; Roche Diagnostics, prod.~no.~ 10837059001), sodium chloride (NaCl; Fisher Chemicals), guanidine hydrochloride (GuHCl; Sigma, prod.~no.~ G4505) and sodium acetate (NaAc; Merck, prod.~no.~ 1.06268) were used without further purification.
Lysozyme purchased from Roche Diagnostics was used for the light scattering experiments and for some cloud-point measurements, which led to consistent findings with the experiments using lysozyme purchased from Sigma Aldrich.

Ultrapure water with a minimum resistivity of 18~M$\Omega$cm was used to prepare buffer and salt stock solutions.
They were filtered several times very thoroughly (nylon membrane, pore size $0.2~\upmu\mathrm{m}$) in order to remove dust particles.
The protein powder was dissolved in a 50~mM NaAc buffer solution.
The solution $p$H was adjusted to $p$H 4.5 by adding small amounts of hydrochloric acid. 
At this $p$H value, lysozyme carries a positive net charge $Q = + 11.4~e$,\cite{Tanford1972} where $e$ is the elementary charge. 
Solutions with an initial protein concentration $c \approx 40-70$~mg/mL were passed several times through an Acrodisc syringe filter with low protein binding (pore size $0.1~\upmu\mathrm{m}$; Pall, prod.~no.~ 4611) in order to remove impurities and undissolved proteins. 
Then, the protein solution was concentrated by a factor of $4-7$ using a stirred ultrafiltration cell (Amicon, Millipore, prod.~no.~ 5121) with an Omega 10 k membrane disc filter (Pall, prod.~no.~ OM010025). 
The retentate was used as concentrated protein stock solution.
Its protein concentration was determined by UV absorption spectroscopy or refractometry.\cite{Platten2015b} 
Protein concentrations $c$ are related to the protein volume fraction $\phi = c \, v_\text{p}$, where $v_\text{p} = 0.740$~mL/g is the specific volume of lysozyme, as inferred from densitometry.\cite{Platten2015b} 
Sample preparation was performed at room temperature $(21\pm2)~^\circ\mathrm{C}$.
Samples with protein volume fractions up to $\phi \approx 0.1$ were prepared by mixing appropriate amounts of buffer, protein and salt stock solutions.
Samples with cloud-points close to or above room temperature were slightly preheated to avoid (partial) phase separation upon mixing.

\subsection{Macroscopic state diagram: determination of the binodal and arrest line}
The metastable gas--liquid coexistence curves (binodals) were determined by cloud-point temperature (CPT) measurements. 
Samples were analyzed immediately after preparation before protein crystals were formed.\cite{Sedgwick2005,Hentschel2021} 
Samples with a typical volume of 0.1~mL were filled into thoroughly cleaned NMR tubes with 5~mm diameter, sealed, and placed into a thermostated water bath at a temperature well above the CPT. 
For an accurate determination of the temperature of the sample solution, a wire thermometer (Dostmann electronic P650, Wertheim, Germany) with a precision of 0.01~$^\circ$C was mounted in a separate, but closely placed NMR tube filled with 0.1~mL water. 
Then, the temperature of the water bath was lowered at a low cooling rate (typically less than 0.3~$^\circ$C/min), during which the sample solution was visually observed. 
The CPT was identified by the sample becoming turbid. 
Further details have been given previously.\cite{Platten2015b}
For moderately concentrated samples (up to $\phi \approx 0.1$), solution compositions have been prepared and analyzed at least three times.

To study samples with $\phi \gtrsim 0.1$, solutions typically with $\phi \approx 0.1$ ($c=120~\mathrm{mg/mL}$) and dedicated salt composition were prepared. Upon a shallow quench (about $1-3$~K) below their cloud points, the sample turned cloudy indicating phase separation.
After quenching, samples (typical volume $0.3~\mathrm{mL}$) 
were placed in a centrifuge (Hermle Z323K) that was set to the respective final temperature and operated at 3,350~g for 15~min.
After centrifugation, the samples were composed of two, transparent or weakly turbid liquid phases.
The protein-rich bottom and the protein-poor top phase were macroscopically separated by a sharp interface.
The dilute phase was removed and its volume and concentration determined. 
Then, the concentration of the dense liquid phase was inferred and its CPT measured.

In order to locate the arrest line, solutions were quenched deeply into the unstable region of the state diagram and subsequently centrifuged to separate solid-like and liquid phases, as in previous studies.\cite{Cardinaux2007,Gibaud2011,Bucciarelli2015}
Solutions with $\phi \approx 0.1$ and dedicated solution conditions were rapidly and deeply quenched (about $5-17$~K) below their cloud points.
At these quench depths, arrested spinodal decomposition occurred and the samples exhibited solid-like behavior on a macroscopic level. 
After quenching, the samples were centrifuged (Hermle Z323K) at 3,350~g for 30~min, such that the structures of the dilute phase yielded and the sample phase separated macroscopically into a dilute liquid and a dense arrested phase.
The dense phase did not fluidize as its meniscus did not tilt. 
The concentration and volume of the dilute phase were measured, based on which the concentration of the dense phase was estimated.
We carefully varied the acceleration and the centrifugation time to make sure that the spinodal network is fully broken and the arrested phase sedimented.
Statistical uncertainties in concentration are indicated for compositions that have been prepared and analyzed at least three times.

\subsection{Microscopic investigation of the condensed states}

For selected samples, the microscopic morphologies of the condensed protein states
as well as the phase separation kinetics (domain formation, coarsening and arrest) were monitored by conventional optical microscopy.
Sample solutions were prepared at a temperature above the cloud-point, typically at $30~^\circ\mathrm{C}$, and filled into a capillary.\cite{Jenkins2008}
The capillary was mounted onto a home-built temperature-controlled microscope stage equipped with a thermoelectric cooler.
The samples were quenched with about $0.5~\mathrm{K/s}$ to specific temperatures and observed using an inverted brightfield microscope (Nikon Eclipse Ti-2) equipped with a $10\times$ plan-fluor objective (Nikon) and a CMOS camera (Allied Vision, Mako U-130B, $512 \times 512~\mathrm{px}^2$) for at least $2~\mathrm{h}$. 
The pixel pitch was $0.485~\upmu\mathrm{m}/\mathrm{px}$.

\subsection{Light scattering: determination of the second virial coefficient}
The second virial coefficient $B_2$ was determined using static light scattering (SLS).\cite{George1994,Muschol1995,Guo1999} 
The experiments were performed with a 3D light scattering apparatus (LS Instruments AG) using only one beam with a wavelength $\lambda =632.8$~nm.
Dilute samples with $3~\text{mg/mL} \leq c \leq 21~\text{mg/mL}$ ($0.002 \leq \phi \leq 0.010$) were investigated at selected temperatures $10.0~^\circ\mathrm{C} \leq T \leq 40.0~^\circ\mathrm{C}$. 
The samples were filled into thoroughly cleaned cylindrical glass cuvettes (diameter 10~mm), centrifuged (Hettich Rotofix 32A) at least 30~min at typically 2,500~g prior to the measurements.
They were then very carefully placed into the temperature-controlled vat of the instrument filled with decalin.
The time-averaged scattered intensity was recorded for typically 100~s with count rates of $5-220$~kHz and about $4$~kHz for protein and buffer solutions, respectively.
In order to check sample quality, dynamic light scattering (DLS) experiments were simultaneously performed on the same samples. 
Samples with indications of aggregates or dust particles were discarded.
For each concentration, at least four independent samples were prepared and measured successfully.

The absolute scattering intensity, i.e.~the excess Rayleigh ratio $R$, varies with protein concentration $c$ and temperature $T$. 
It was determined at $\lambda =632.8$~nm using toluene as a reference according to
\begin{eqnarray}
R(c,T) = \frac{\left\langle I_\text{p} (c,T) \right\rangle - \left\langle I_\text{s} (T) \right\rangle}{\left\langle I_\text{t} (T) \right\rangle - \left\langle I_\text{dc} \right\rangle} \, \frac{n(c,T)^2}{n_\text{t}(T)^2} \, R_\text{t}(T)
\end{eqnarray}
with the time-averaged scattered intensities of sample, solvent, toluene, and background (blocked beam), $\left\langle I_\text{p} \right\rangle$, $\left\langle I_\text{s} \right\rangle$, $\left\langle I_\text{t} \right\rangle$, and $\left\langle I_\text{dc} \right\rangle$, respectively, 
the refractive indices of the sample and toluene, $n$ and $n_\text{t}$, and
the Rayleigh ratio of toluene at the measurement temperature with $R_\text{t}= 1.40 \times 10^{-5}~\text{cm}^{-1}$ at $T=35^\circ\text{C}$\,\cite{Bender1986,Narayanan2003,Itakura2006}.
The temperature dependence of $R_\text{t}$ was determined from the temperature dependence of the intensity scattered by a toluene sample.\cite{Goegelein2012}

The refractive index of sample solutions $n$ was measured with a temperature-controlled Abbe refractometer (Model 60L/R, Bellingham \& Stanley) operated with a HeNe laser ($\lambda=632.8~\mathrm{nm}$) and the respective temperature of the SLS experiment. Refractive index increments, $\text{d}n/\text{d}c$, were obtained from linear fits to the dependence of $n$ on $c$. 

In one-component solutions, the excess scattering typically contains information on the shape and size of the particles as well as the particle arrangement.
Their contributions are reflected in the form factor $P(Q)$ and the structure factor $S(Q)$, respectively,  
where $Q = (4 \pi n / \lambda) \sin{(\theta/2)}$ is the magnitude of the scattering vector with the scattering angle $\theta$.
Then, the excess Rayleigh ratio $R$ reads:
\begin{eqnarray}\label{eq:rkc}
R(Q) = K\, c \, M \, P(Q) \, S(Q)
\end{eqnarray}
with the averaged molar mass of the particle $M$ and an optical constant $K$ given by
\begin{eqnarray}
K(T) = \frac{4 \pi^2 n_\text{s}(T)^2}{N_\text{A} \lambda^4} \left( \frac{\text{d}n}{\text{d}c}(T) \right)^2 
\end{eqnarray}
with the refractive index of the solvent $n_\text{s}$ and Avogadro's number $N_\text{A}$.

In most of our experiments, $\theta=90^\circ$ and hence $Q\approx 0.018~\text{nm}^{-1}$.
The effective protein diameter $\sigma=3.4~\text{nm}$, as inferred from the molecular size, is small compared to $\lambda$, implying $\sigma \, Q \ll 1$.
In the small-$Q$ limit, lysozyme acts as a point scatterer, i.e.~$P(Q\to 0)= 1$. 
For selected samples, experiments were performed at $30^\circ \leq \theta \leq 150^\circ$.
$R$ was found to be independent of $\theta$ and hence of $Q$, as expected for point scatterers.
In addition to the DLS experiments, this also suggested that large particles, such as impurities or aggregates, were absent.

Moreover, $S(Q\to 0)= \kappa_\mathrm{T} / \kappa_\mathrm{T}^\mathrm{id}$ with the isothermal osmotic compressibility of the sample and an ideal solution, $\kappa_\mathrm{T}$ and $\kappa_\mathrm{T}^\mathrm{id}$, respectively.\cite{Hansen2006}
In the limit of low $c$, $\kappa_\mathrm{T}^{-1}$ can be expressed in terms of a virial expansion.
Then, Eq.~(\ref{eq:rkc}) can be rearranged and simplified, such that the second virial coefficient $B_2$ can be determined in a model-independent manner from $R(c)$ via:
\begin{eqnarray}\label{eq:kcp}
\frac{K c}{R(c)} = \frac{1}{M^{(0)}} \left(1 + \frac{2 N_\text{A} B_2}{M^{(1)}} \, c \right) \, ,
\end{eqnarray}
where $M^{(0)}$ and $M^{(1)}$ represent the molar mass. 
We take $M^{(1)} = 14 320~\text{g/mol}$, so that uncertainties in the apparent molar mass $M^{(0)}$ do not affect the determination of $B_2$.\cite{Goegelein2012} 
For $M^{(0)}$, we obtain values in the range $0.8 \lesssim M^{(0)}/M^{(1)} \lesssim 1.0$.
The differences between $M^{(0)}$ and $M^{(1)}$ might be ascribed to the low scattered intensities of the protein solutions, solvents and toluene and uncertainties in $\text{d}n/\text{d}c$ and $R_\text{t}$.\cite{Goegelein2012} 
They might also reflect protein--water and protein--cosolute interactions.\cite{Blanco2011}

\subsection{Non-equilibrium self-consistent generalized
Langevin equation (NE-SCGLE) theory}

The non-equilibrium self-consistent generalized Langevin equation (NE-SCGLE) theory of irreversible processes in liquids was introduced and applied before.\cite{Medina2009,Ramirez2010,Ramirez2010a,Sanchez2013,Sanchez2014}
We consider the hard-sphere plus attractive Yukawa (HSAY) model,
defined by the pair potential: 
\begin{eqnarray}\label{eq:yu}
U_\text{HCAY}(r) = \begin{cases}\infty, \quad & r < \sigma; \\-\epsilon \frac{\exp{[-z(r/\sigma-1)]}}{r/\sigma} , \quad & r \geq \sigma. \end{cases} 
\end{eqnarray}
For given $\sigma$, $\epsilon$ and $z$, the state space of this system is spanned by the dimensionless number
density $\rho\sigma^3$ and temperature $k_\text{B}T/\epsilon$, denoted simply as $\rho^\star$ and $T^\star$; we also refer to the hard-sphere volume fraction $\phi=\pi \rho^\star/6$. 
The dimensionless time $D_0 t /\sigma^2$ (with the colloidal short-time self-diffusion coefficient $D_0$\cite{Lopez2012}) is also denoted simply as $t^\star$.

This model, with $z = 2$, was employed in Ref. \citenum{Olais-Govea2015} to analyze the asymptotic stationary
solutions of the most fundamental equation of the NE-SCGLE theory (Eq. (\ref{eq:ne}) below), to
predict the main features of the dynamic arrest diagram. 
As explained in Refs. \citenum{Sanchez2013,Olais-Govea2015},
the NE-SCGLE theory describes the spontaneous response of an instantaneously quenched
liquid in terms of the non-equilibrium static structure factor $S(k; t)$, whose time-evolution
equation for $t > 0$ reads
\begin{eqnarray}\label{eq:ne}
\frac{\partial S(k;t)}{\partial t} = -2k^2 D_0 b(t) n \mathcal{E}_f(k)
\left[S(k;t) -1/n\mathcal{E}_f(k)\right]
\,.
\end{eqnarray}
The details of these calculations are identical to those explained in Ref. \citenum{Olais-Govea2015}.

\section{\label{sec:res}Results and Discussion}

Protein solutions are studied at almost molar salt concentrations, such that electrostatic repulsions are largely screened and the interactions are expected to be dominated by short-range attractions.
For three additive compositions, detailed macroscopic non-equilibrium state diagrams, including the liquid--liquid phase coexistence line and the dynamical arrest line, are established and compared with microscopic observations.
Then, for a broad range of additive compositions and temperatures,
the dependence of the excess scattered intensity of protein solutions on the protein concentration is experimentally determined and second virial coefficients are inferred. 
For each additive composition, the LLPS critical temperature $T_\text{c}$ is used to rescale the state boundaries as well as the temperature dependence of $b_2$.
Finally, the experimental results are compared with calculations based on the NE-SCGLE theory.

 \subsection{Experimental state diagram}

\begin{figure}
	\centering
	\includegraphics[width=0.55\textwidth]{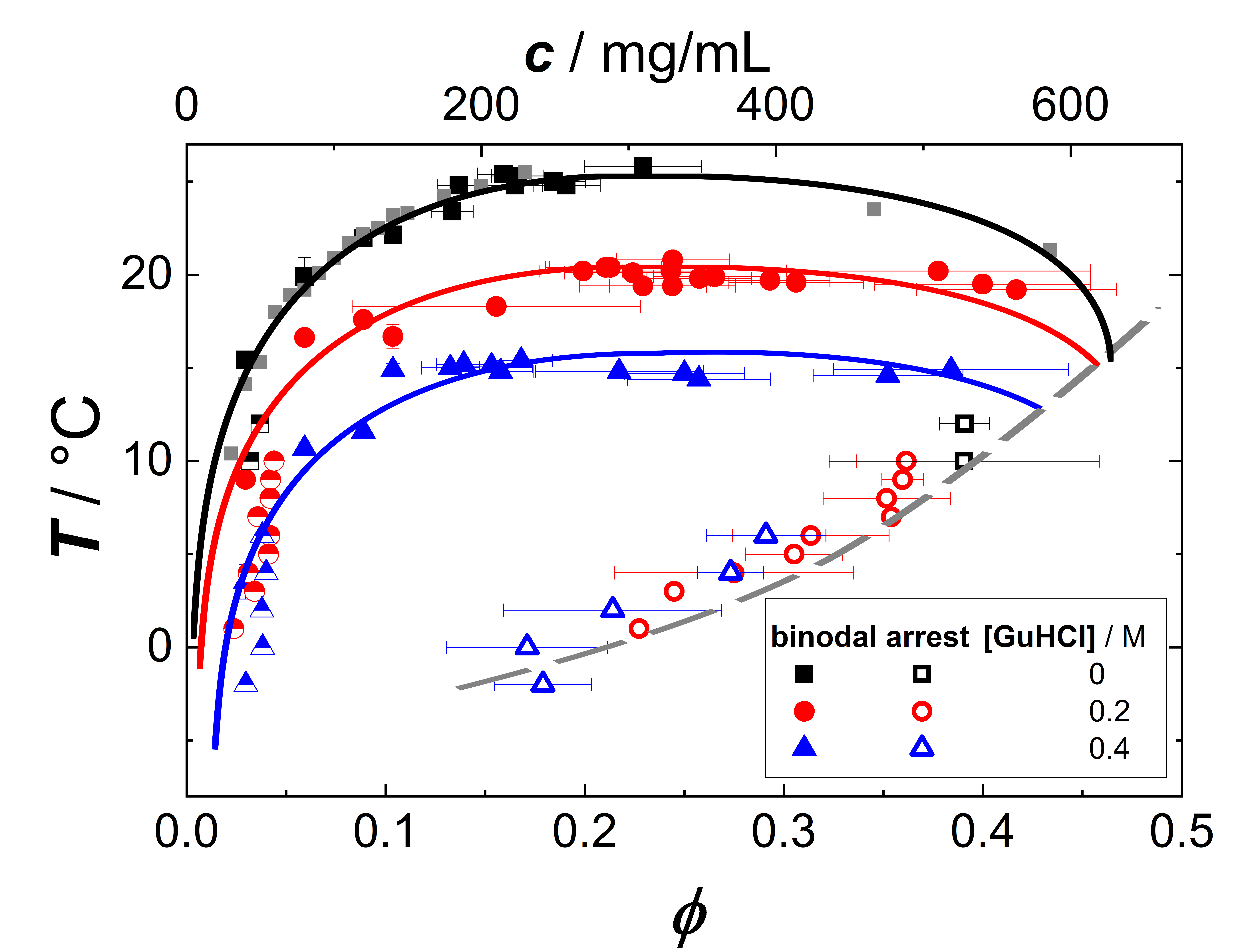}
	\caption{ Experimentally determined state diagrams (temperature $T$ vs.~volume fraction $\phi$ (bottom) or protein concentration
$c$ (top)) of protein solutions (lysozyme, $p$H 4.5) containing 0.9~M NaCl and different amounts of GuHCl as indicated.
Liquid--liquid phase separation (LLPS) and dynamical arrest lines are shown as filled and open
symbols, respectively. 
Lines are guides to the eye.
Half-filled symbols denote the dilute phase, which coexists with the arrested state.
For completeness, LLPS data from ref. \citenum{Goegelein2012} (grey squares) and a few LLPS data points obtained at low $\phi$ (same symbols as current data set) from ref. \citenum{Platten2015b} have been included.  
}
\label{fig:2}
\end{figure}

In this section, experimental non-equilibrium state diagrams of lysozyme solutions dominated by short-range attractions are presented.
Fig.~\ref{fig:2} shows the metastable liquid--liquid phase coexistence (or binodals) and dynamical arrest lines for three different additive compositions in the temperature $T$ vs. volume fraction $\phi$ (or protein concentration $c$) plane.

For solutions containing only NaCl, the binodal (filled black squares) has the following shape:
In the low-density (gas) branch, the phase coexistence (cloud-point) temperature $T_\text{LLPS}$ increases strongly with $\phi$, but saturates at moderately high $\phi$ ($\phi \gtrsim 0.15$).
Close to the critical point and in the high-density (liquid) branch, the binodals appear to be almost flat.
Moreover, the data quantitatively agree with previous measurements (filled grey squares)\cite{Goegelein2012}. 
For solutions containing GuHCl in addition, the shape of the binodals (filled red circles and filled blue triangles) appears to be similar.
Such strongly asymmetric shapes have been observed before\cite{Cardinaux2007,Gibaud2011,Goegelein2012,Platten2015}
and might reflect the patchiness of the protein--protein interactions\cite{Goegelein2008,Kastelic2015}.
While the shape appears to be unaltered by the GuHCl content,
the location of the binodal is systematically affected.
According to the previously observed almost linear decrease of cloud-points with [GuHCl] at fixed low $\phi$,\cite{Platten2015b} 
Fig.~\ref{fig:2} shows that the whole binodal shifts to lower temperatures upon addition of GuHCl, indicating reduced net attractions.

The dynamical arrest lines (open symbols) reach deeply inside the demixing region.
They increase with $\phi$ and cross the binodals at high $\phi$.
In contrast to the binodals which depend on the additive composition, the dynamical arrest lines of the three systems seem to coincide in the $T$ vs.\,\,$\phi$ plane irrespective of the guanidine content.
A similar collapse of arrest lines has been observed for lysozyme solutions with mixed interactions ($p$H 7.8 and various amounts of NaCl).\cite{Gibaud2011} 

The low-density fluid phase that coexists with the arrested state is shown in addition (half-filled symbols).
For fixed $T$, its volume fraction is slightly larger than that of the gas-branch of the binodal, which might indicate an effect of the arrest on the phase separation.\cite{Kroy2004}

\begin{figure}
	\centering
	\includegraphics[width=0.55\textwidth]{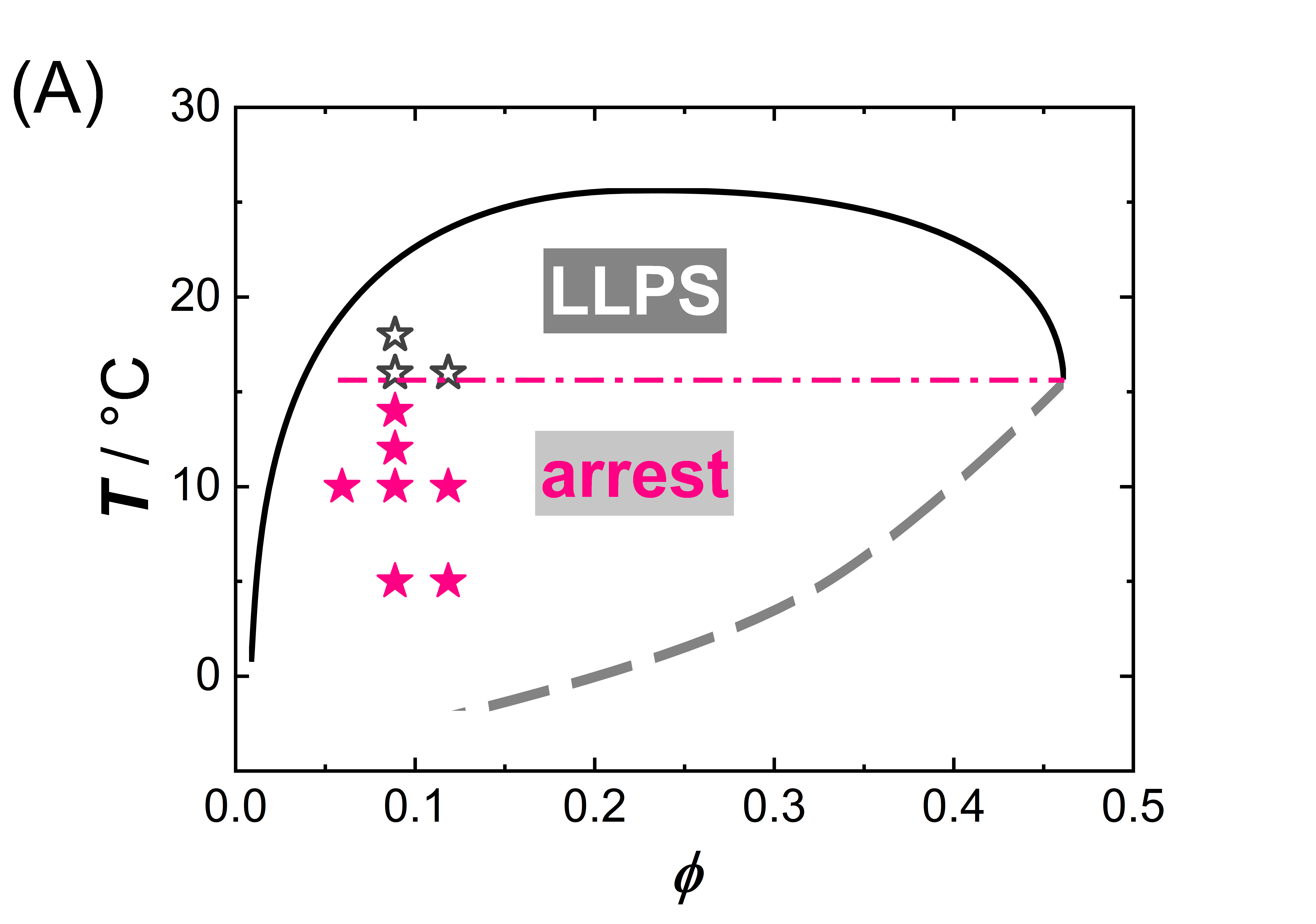}
	\includegraphics[width=0.85\textwidth]{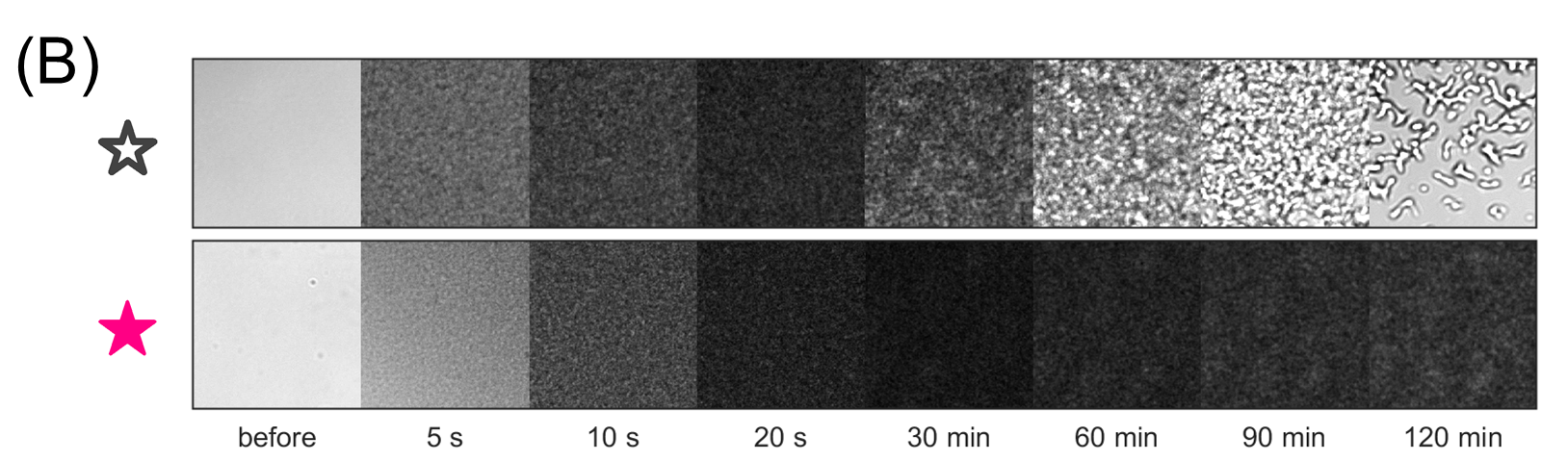}
	\caption{ (A) Experimentally determined state diagram (temperature $T$ vs.\,volume fraction $\phi$ (lysozyme, $p$H 4.5, 0.9~M NaCl); interpolated
liquid--liquid phase separation (LLPS) binodal line (solid) and dynamical arrest line (dashed) as in Fig.~\ref{fig:2}.
Their intersection point determines the arrest tie line (dash-dotted).
The conditions probed by optical microscopy are indicated (stars): phase-separating (open) and dynamically arrested samples (filled magenta). 
(B) Optical micrographs (image size 248 x 248 $\upmu$m$^2$) illustrating the phase transition kinetics of samples undergoing spinodal decomposition (top) and arrested spinodal decomposition (bottom); in this case, $\phi=0.09$ at $T=16~^\circ\mathrm{C}$ (top) and $T=12~^\circ\mathrm{C}$ (bottom), respectively.
}
\label{fig:new}
\end{figure}

In Fig.~\ref{fig:new}(A), the binodal of solutions only containing NaCl as well as the arrest lines are replotted. 
The temperature $T_\text{a}\approx 15~^\circ\mathrm{C}$, where the arrest line crosses the binodal, defines the arrest tie line (dash-dotted).
Thus, below the LLPS binodal, distinct regions of the state diagram can be identified:
For shallow quenches below the binodal, i.e.\, $T_\text{LLPS}>T>T_\text{a}$, macroscopic phase separation procedes via spinodal decomposition. 
For deeper quenches, i.e.\, $T_\text{LLPS}>T_\text{a}>T$, the interplay between phase separation and dynamical arrest can induce the formation of gel-like states via arrested spinodal decomposition, in which a dilute fluid coexists with a dense arrested phase.

In order to obtain a more comprehensive picture of the different states attained by protein solutions dominated by short-range attractions, the phase transition kinetics of selected samples (marked as stars in  Fig.~\ref{fig:new}(A)) were monitored by optical microscopy. 
Indeed, samples quenched above and below $T_\text{a}$ exhibit qualitatively different phase transition kinetics, as indicated by open and filled stars in Fig.~\ref{fig:new}(A).
Fig.~\ref{fig:new}(B) shows typical examples of the temporal evolution of micrographs of samples with $\phi=0.09$ at $T=16~^\circ\mathrm{C}$ (top) and $T=12~^\circ\mathrm{C}$ (bottom), respectively. 
For the shallow quench (top), the micrographs initially darken as the sample becomes turbid and, at a later stage, they show domain formation and coarsening kinetics as a feature of spinodal decomposition.
While, for deeper quenches (bottom), the initial opacification is similar, the micrographs at later stages do not evolve with time. 
The network structure formed via spinodal decomposition appears to be dynamically frozen.  
Hence, the different phase transition kinetics observed by optical microscopy confirm the location of $T_\text{a}$, estimated from the macroscopic state diagram, as well as the suggested mechanisms in line with previous experiments on protein solutions with mixed interactions\cite{Cardinaux2007,Gibaud2009}.

\begin{figure}
	\centering
	\includegraphics[width=0.55\textwidth]{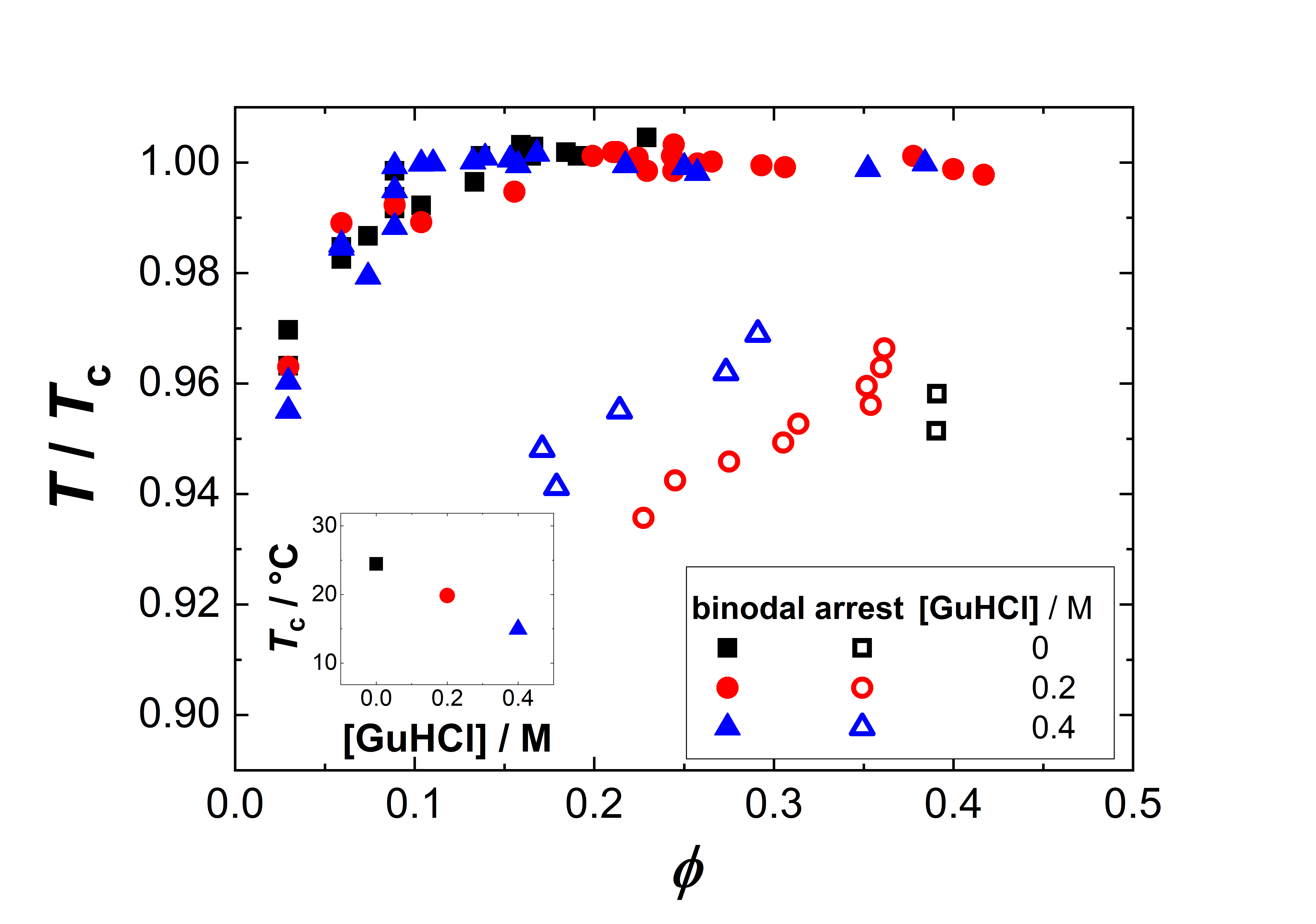}
	\caption{ 
	State diagrams as in Fig.~\ref{fig:2}, but with the temperature axis normalized by the
respective critical temperature $T_\text{c}$ (shown as an inset).
}
\label{fig:ttc}
\end{figure}

In order to further analyze the state diagrams (as well as the underlying interactions), one point of each binodal is particularly important: its maximum with the critical temperature $T_\text{c}$.\cite{Stanley1971}
Estimates of $T_\text{c}$ are displayed in the inset of Fig.~\ref{fig:ttc}.
It decreases approximately linearly with GuHCl. 

Now, for each additive composition, $T_\text{c}$ can be used to rescale the temperature dependence of the state diagram.
Fig.~\ref{fig:ttc} shows the data presented in Fig.~\ref{fig:2} with a temperature axis normalized by the critical temperature of the specific binodals.
Then, 
the different arrest lines are separated from each other, while
the three binodals collapse onto one another, further indicating the similarity of their shapes.
This implies that, just as $T_\text{c}$, the whole phase coexistence curve decreases almost linearly with GuHCl.

\subsection{Protein--protein interactions close to phase separation}

Light scattering experiments were used to characterize the protein--protein interactions.
The interactions are modulated by the addition of sodium chloride (NaCl) and guanidine chloride (GuHCl), focussing on solution conditions (additive compositions and temperatures) relevant for phase separation (cf.~Fig.~\ref{fig:2}), but on dilute protein concentrations, such that $b_2$ values can be determined.

\begin{figure}
	\centering
	\includegraphics[width=0.48\textwidth]{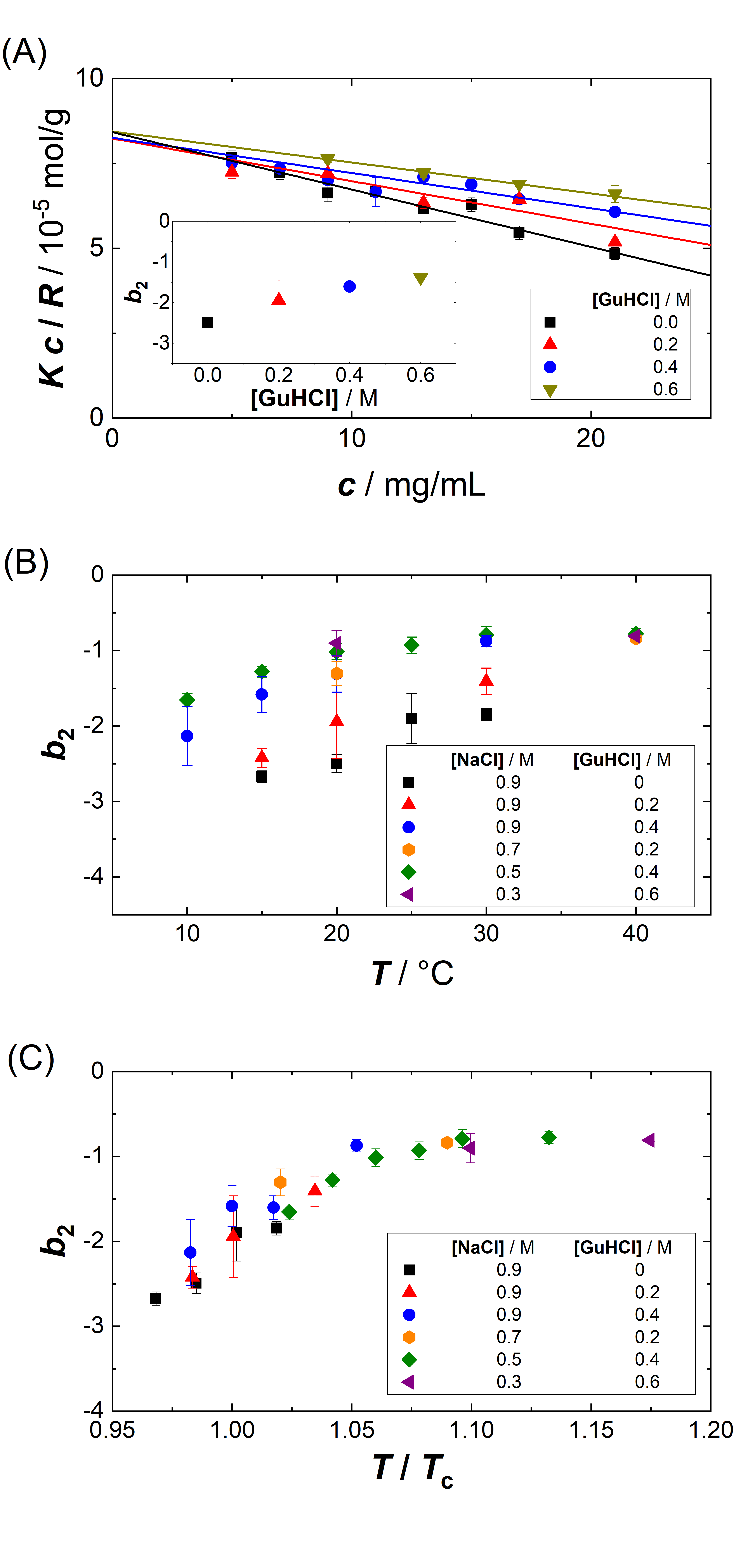}
	\caption{ 	Interactions in protein solutions (lysozyme, $p$H 4.5, with indicated amounts of NaCl and GuHCl) as inferred from static light scattering:
	(A) Debye plots: Inverse of the normalized scattered intensity $K c / R$ vs. concentration $c$ of protein solutions with 0.9~M NaCl and various GuHCl concentrations as indicated at $T=20.0~^\circ\mathrm{C}$ (symbols) and linear fits (lines) according to Eq.~(\ref{eq:kcp}).   
	(Inset) Dependence of the normalized second virial coefficient $b_2=B_2/B_2^\mathrm{HS}$, as inferred from the slope of the linear fits, on the GuHCl concentration. 
	(B) Temperature $T$ dependence of $b_2$ of protein solutions, as obtained from static light scattering, with various salt concentrations as indicated.
	(C) Normalized second virial coefficient $b_2$ as a function of
$T / T_\text{c}$ for the indicated solution conditions (with $T_\text{c}$ values provided in the inset of Fig.~\ref{fig:ttc} or, for additional data sets, in ref.~\citenum{Hansen2016}). 
}
\label{fig:1}
\end{figure}

Fig.~\ref{fig:1}(A) displays exemplary Debye plots: 
The inverse of the normalized excess scattered intensity, $K c / R$, is shown as a function of the protein concentration $c$ for four different additive compositions (fixed NaCl, but various GuHCl concentrations, indicated by different symbols) at a fixed temperature $T=20.0~^\circ\mathrm{C}$.
The $c$ dependence of the data is described by linear fits (lines).
The intercept and thus the measured molecular weight are essentially independent of the additive content.
The slope of all fit curves is negative, indicating net attractions between the protein molecules and negative $b_2$ values.
However, with increasing GuHCl content, the magnitude of the slope decreases, thus reflecting weakening net attractions and a decreasing magnitude of $b_2$ (inset).
Similarly, at fixed $T$ and $\phi$, the distance to the LLPS binodal increases with [GuHCl] (cf.~Fig.~\ref{fig:2}).

On a molecular level, protein--protein interactions are likely affected by patchiness, asymmetric shapes, and charge patterns; additives, like guanidine ions\cite{Liu2005}, can preferentially bind to protein surfaces and alter their hydration and hydrogen bonding properties.
Nevertheless, in previous works,\cite{Goegelein2012,Platten2016,Hansen2021b} 
a coarse-grained macroscopic picture based on Derjaguin-Landau-Verwey-Overbeek (DLVO) theory using an appropriate Hamaker constant was sufficient to capture their effect on $b_2$ of protein solutions.

Similar experiments as shown in Fig.~\ref{fig:1}(A) were performed for a broad range of additive compositions and at different temperatures. 
The resulting $b_2$ values are shown in Fig.~\ref{fig:1}(B) as a function of $T$.
All $b_2$ values are negative, indicating net attractions, but the magnitude of $b_2$ systematically varies with solution composition and temperature.
For a given solution condition, $b_2$ increases (becomes less negative) with $T$, i.e.\,\,the net attractions become less pronounced upon increasing $T$ for these systems.
At fixed $T$, the magnitude of $b_2$ can be increased and decreased by the addition of NaCl and GuHCl, respectively.
Correspondingly, these additives are known to shift the LLPS phase boundary of lysozyme to higher and lower temperatures, respectively.\cite{Platten2015b}

Fig.~\ref{fig:1}(C) shows the data presented in Fig.~\ref{fig:1}(B) with the temperature axis normalized by the respective critical temperature of the binodals.
(For solution conditions not shown in Fig.~\ref{fig:2}(A) $T_\text{c}$ values were taken from the literature.\cite{Hansen2016})
Remarkably, the $b_2$ data collapse onto a master curve.
(Note that the present data set also agrees with previous results\cite{Goegelein2012,Hansen2021b,Hansen2022} by light and X-ray scattering as shown in the Appendix.)
This implies that, 
in the vicinity of the binodal, the various systems studied have the same pair interaction strength, which is governed by only one parameter, the temperature relative to the critical temperature.
Similar collapses (despite slightly different values of $b_2(T_\text{c})$) have been previously found for lysozyme\cite{Platten2015,Hansen2021b} and $\gamma$B-crystallin\cite{Bucciarelli2016}.
Our value of $b_2(T_\text{c})\approx -1.9$ is close to the one proposed by Vliegenthart and Lekkerkerker \cite{Vliegenthart2000} and supported by simulations\cite{Valadez-Perez2012} for many systems dominated by short-range attractions, $b_2(T_\text{c})\approx-1.5$. 
The difference might be ascribed to the effect of repulsions on the effective particle size, as noted previously\cite{Noro2000,Platten2015}.

\begin{figure}
	\centering
	\includegraphics[width=0.55\textwidth]{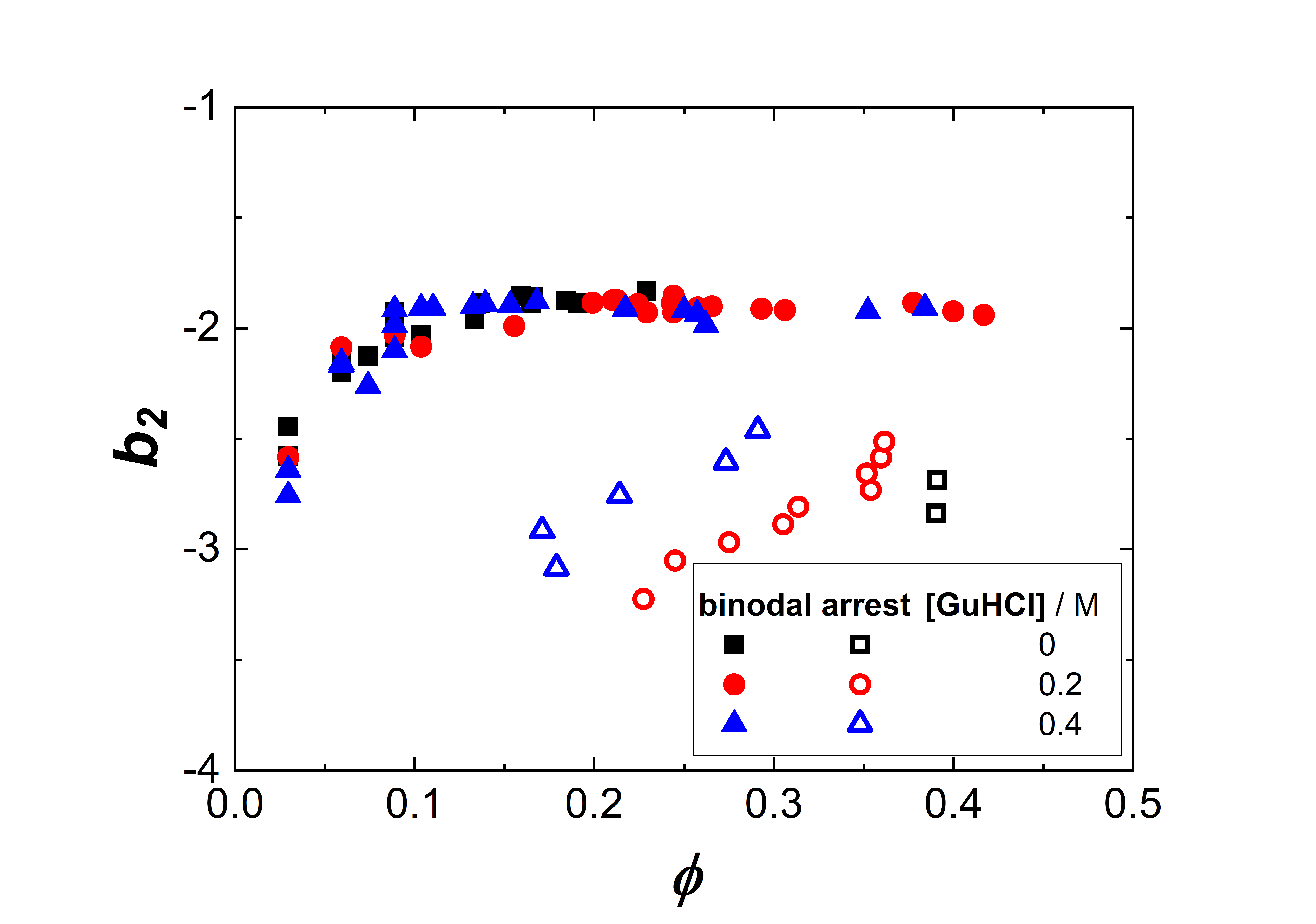}
	\caption{ 
	 State diagrams
as in Fig.~\ref{fig:2}, but with $T$ being replaced by $b_2$. 
}
\label{fig:b2p}
\end{figure}

The data shown in Fig.~\ref{fig:1}(C) can also be used to replace the normalized temperature axis in Fig.~\ref{fig:2} (or Fig.~\ref{fig:ttc}) with $b_2$.
Fig.~\ref{fig:b2p} shows the respective state diagrams with the ordinate reflecting the net strength of the attractions.
Again, as in Fig.~\ref{fig:ttc}, the binodals nicely collapse onto one another, but the arrest lines separate from each other.
In line with previous results,\cite{Gibaud2011,Platten2015} these findings indicate that the binodals can be rescaled by an integral measure of the interaction potential and do not depend on the details of the interaction potential.
Thus, our experimental data support the applicability of the extended law of corresponding states to the binodals of protein solutions.\cite{Noro2000,Platten2015}  
(Note that the shape of the binodal might change if the repulsive interactions are altered.\cite{Platten2015})
However, for the arrest lines, the details of the short-range attractions seem to be important, in particular the attraction strength which might depend on the (absolute) temperature.
This difference between the binodal and arrest lines might be related to the fact that in a fluid the particles sample all distances, whereas in the arrested state only the shortest distances are relevant and thus the strength rather than the range of the potential is crucial.

For colloids with short-ranged attractions induced by polymers, gelation has been ascribed to spinodal decomposition inducing the formation of a space-spanning cluster.\cite{Lu2008}
In this scenario, a universal state diagram of the arrest lines if scaled by $B_2$ would be expected.
However, this is in contrast to our findings, where the arrest lines
extend far into the unstable region below the coexistence curve, as previously observed for proteins with mixed (short-range attractive and long-range repulsive) potentials\cite{Cardinaux2007,Bucciarelli2015}. 
It has been speculated that these different routes to gelation might be related to the role of directional interactions.\cite{DaVela2020}

\subsection{Theoretical state diagram}

Despite the inherent complexity of protein--protein interactions, e.g., due to their directionality, coarse-grained models have proven helpful to rationalize protein phase behavior, such as an (isotropic) DLVO model.\cite{Muschol1995,Poon2000,Pellicane2012,Goegelein2012,Hansen2021b} 
Furthermore, when repulsions are screened and attractions dominate, the spherically symmetric hard-core attractive Yukawa (HCAY) fluid can provide a simplified description of the interactions.\cite{Rosenbaum1999,Foffi2002,Orea2010,Valadez-Perez2012,Gazzillo2013}

The non-equilibrium self-consistent Langevin equation (NE-SCGLE) theory of irreversible relaxation has been applied to HCAY systems with an attraction range smaller than the particle size.\cite{Olais-Govea2015} The main findings were:
At high temperatures and densities, compressing or cooling this liquid drives it through the transition to
non-equilibrium repulsive glasses.\cite{Sanchez2013} 
At intermediate and low densities and temperatures the NE-SCGLE theory predicts that (i) the spinodal line is a frontier between
equilibrium and non-ergodic states, (ii) that the arrest line penetrates inside the
spinodal region, (iii) that there exists a crossover temperature, above which the unstable
homogeneous liquid fully phase separates, and below which it forms gels.

\begin{figure}
	\centering
	\includegraphics[width=0.55\textwidth]{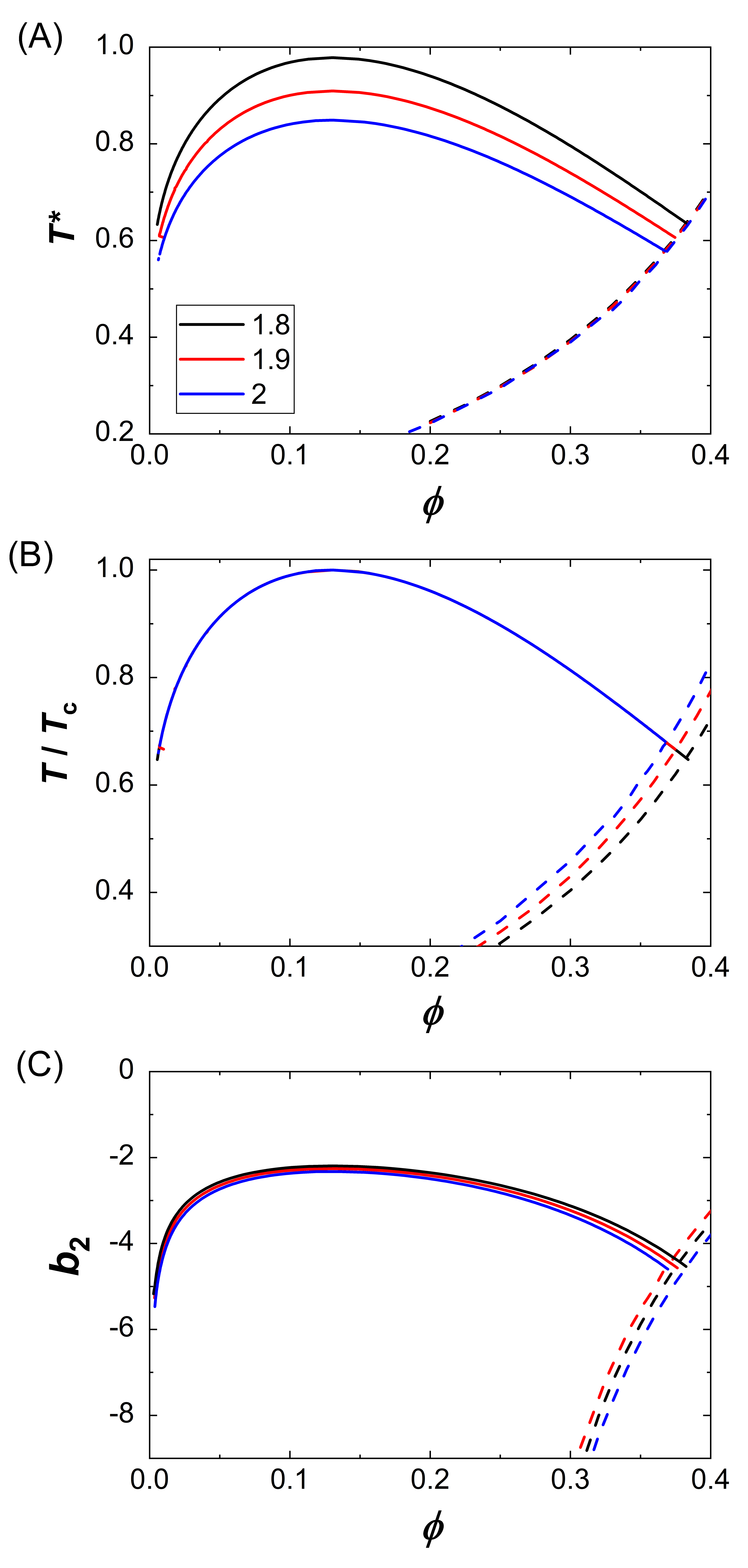}
	\caption{ 
	Theoretical results of the binodal (solid lines) and dynamical arrest line (dash-dotted
lines), obtained as explained in Ref.~\citenum{Olais-Govea2015}, for different values of the range $z$ ($z = 1.8$, $1.9$, and $2$) of the hard-core attractive
Yukawa system (Eq.~(\ref{eq:yu})) (as indicated) with (A) normalized temperature $T^\star$, (B) temperature normalized by the critical temperature $T/T_\text{c}$ 
and (C) second virial coefficient $b_2$ as the vertical axis, respectively. 
}
\label{fig:3}
\end{figure}

This scenario is in line with the present experimental findings as shown in Fig.~\ref{fig:new} (although the effective $z$ of the experiments might be larger) as well as previous work on protein solutions with mixed interactions\cite{Cardinaux2007,Gibaud2011}.
In order to further understand the effects of the additive on the binodal and on the arrest line, further theoretical analysis was carried out here. 
In the experiments, the added electrolyte alters the range and strength of the attractive pair potential. 
Since the
strength is expected to simply renormalize the temperature (as suggested by the universal $T/T_\text{c}$ dependence of $b_2$ in Fig.~\ref{fig:1}(C)), it is conceivable 
that the three additive compositions reported can be modelled as three different values of the range of attractions.

To see if a relation can be established between the scenario predicted by the NE-SCGLE theory and the experimentally observed results, the binodal (instead of the spinodal) line (solid lines) together with the arrest line (dashed lines) are determined for the different values of $z$. 
Due to the complex nature of the protein--protein interactions (compared to the simple model potential) and the approximations made to obtain an analytical description of the model structure factor, we only aim at a qualitative comparison.
For the sake of simplicity, the values of $z$ are varied only in a narrow range close to the previously used value.
The results are shown in Fig.~\ref{fig:3}.
In Fig.~\ref{fig:3}(A), the trends with $z$ exhibited by the three theoretical results plotted in this manner
(using temperature in the vertical axis) seem to coincide with the experimental scenario:
the binodal lines differ among them, but the dynamic arrest lines below the binodal coincide.
The same theoretical results can now be plotted with the temperature normalized by the critical temperature and 
using in the vertical axis $b_2$ as shown in Fig.~\ref{fig:3}(B) and (C), respectively.
The three binodal lines cluster together,
while the three arrest lines now disaggregate to a noticeable extent. 
Qualitatively, this is the same trend observed in the experimental data.

\section{\label{sec:con}Conclusion}

In summary,
the phase behavior and interactions of protein solutions were studied.
The interactions were dominated by short-range attractions, as the repulsions were largely screened by the presence of almost molar salt concentration; the attraction range and strength were tuned by additives. 
While the liquid--liquid coexistence curve was systematically shifted depending on the additive concentration, the dynamical arrest lines collapsed onto one another when using an absolute temperature axis.
The second virial coefficient, and hence the integrated strength of the protein--protein interactions, were found to exhibit a universal temperature dependence if the temperature was measured relative to the critical temperature of the binodal.
Hence, the different binodals collapse onto one another if the temperature axis is replaced by $b_2$, supporting the extended law of corresponding states for the binodals of protein solutions.
However, if $b_2$ instead of $T$ is used, the arrest lines do not coincide, but separate from each other. 
This indicates that they are not governed by corresponding-states pair interactions.
Rather, the attraction strength might be important and the attraction range less crucial.
These findings could be rationalized based on the non-equilibrium self-consistent Langevin equation (NE-SCGLE) theory applied to hard-core attractive Yukawa fluids of different attraction range.
This work hence contributes to a better understanding of the interplay between phase separation and dynamical arrest of systems dominated by short-range attractions.

\section*{Acknowledgements}
F.P. acknowledges financial support by the Strategic Research Fund of the Heinrich Heine University (F 2016/1054-5) and the German Research Foundation (PL 869/2-1).

\section*{Appendix}

\begin{figure}
	\centering
	\includegraphics[width=0.55\textwidth]{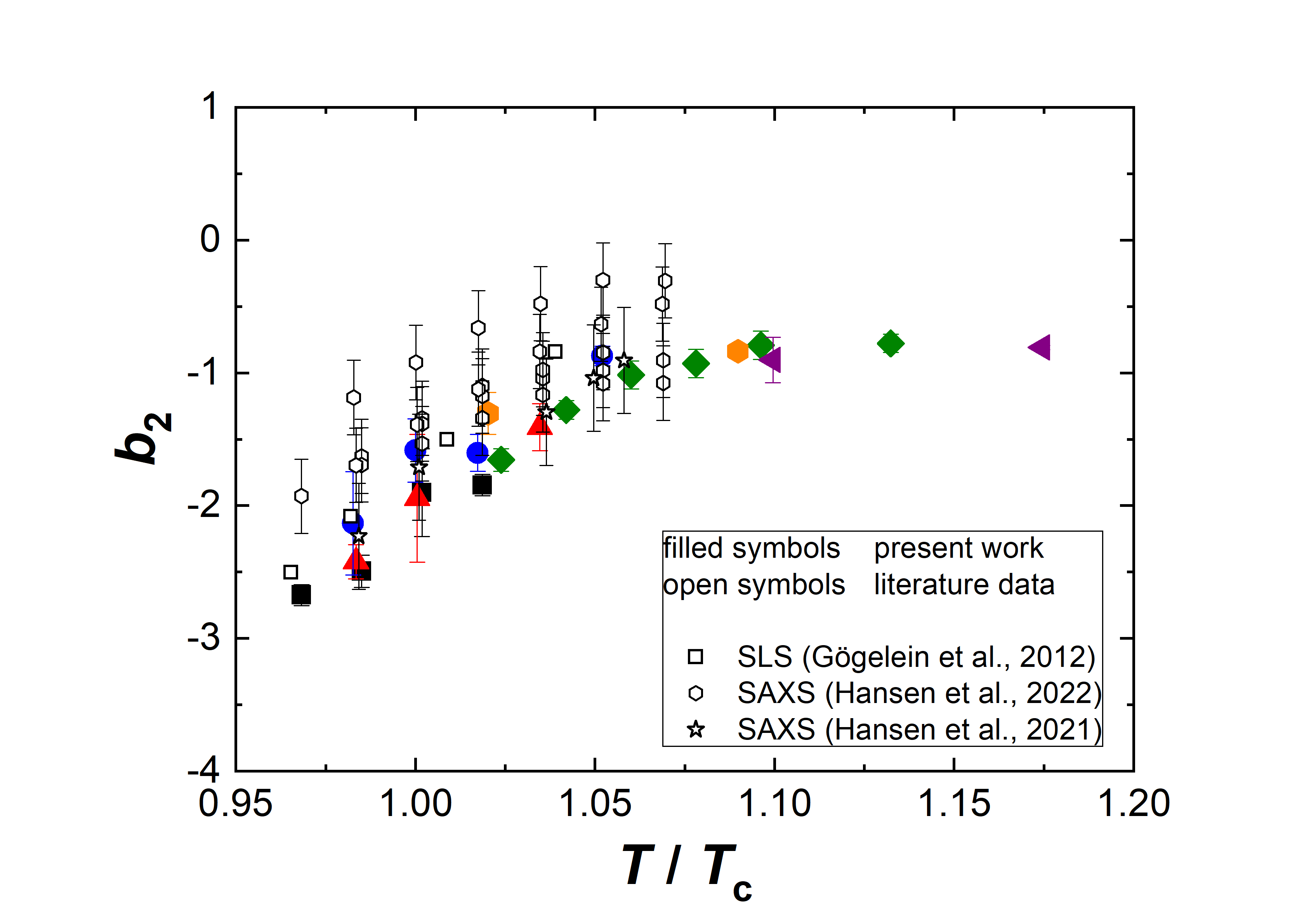}
	\caption{ 
	Normalized second virial coefficient $b_2$ as a function of
$T / T_\text{c}$ for various solution conditions at $p$H 4.5, including the data shown in Fig.~\ref{fig:1}(C) as well as literature data\cite{Goegelein2012,Hansen2021b,Hansen2022}.
}
\label{fig:f}
\end{figure}

In Fig.~\ref{fig:f}, the data of Fig.~\ref{fig:1}(C) are replotted and literature data\cite{Goegelein2012,Hansen2021b,Hansen2022} are added.

\normalem 
   
\bibliography{lorena}  
  
\end{document}